\documentclass[aps,prb,reprint,superscriptaddress,twocolumn, longbibliography]{revtex4-2}

\usepackage{graphicx}
\usepackage{amsmath,amsfonts,amssymb,mathtools}
\usepackage{color}
\usepackage{subfigure}
\usepackage{physics}
\usepackage{dsfont}
\usepackage{hyperref}
\usepackage{booktabs}
\usepackage{url}
\usepackage{leftidx}
\usepackage{textcomp}
\usepackage{gensymb}
\usepackage{threeparttable}
\usepackage{rotating}
\usepackage{esvect}
\usepackage{booktabs}
\usepackage{marvosym}
\usepackage{lmodern}
\usepackage[noend]{algpseudocode}
\usepackage[dvipsnames]{xcolor}

\newcommand{\cd}{c^{\dagger}}

\begin{document}

\title{Time-reversal symmetry breaking from lattice dislocations in superconductors}

\author{Clara N. Brei\o}
\affiliation{Niels Bohr Institute, University of Copenhagen, DK-2200 Copenhagen, Denmark}

\author{Andreas Kreisel}
\affiliation{Niels Bohr Institute, University of Copenhagen, DK-2200 Copenhagen, Denmark}

\author{Mercè Roig}
\affiliation{Niels Bohr Institute, University of Copenhagen, DK-2200 Copenhagen, Denmark}

\author{P. J. Hirschfeld}
\affiliation{Department of Physics, University of Florida, Gainesville, Florida 32611, USA}

\author{Brian M. Andersen}
\affiliation{Niels Bohr Institute, University of Copenhagen, DK-2200 Copenhagen, Denmark}

\date{\today}

\begin{abstract}
Spontaneous generation of time-reversal symmetry breaking in unconventional superconductors is currently a topic of considerable interest.  While chiral superconducting order is often assumed to be the source of such signatures, they can sometimes also arise from nonmagnetic disorder. Here we perform a theoretical study of the impact of dislocations on the superconducting order parameter within a microscopic one-band model which, in the homogeneous case, features either extended $s$-wave, $d$-wave, or $s+id$-wave superconductivity depending on the electron concentration. We find that the dislocations minimize their impact on the superconducting condensate by inducing localized supercurrents pinned by the dislocations, even well outside the $s+id$ regime. We map out the parameter and density dependence of the induced currents. From these results we conclude that quite generically unconventional  superconductors hosting  dislocations tend to break time-reversal symmetry locally. 

\end{abstract}

\maketitle

\section{Introduction}

A growing class of unconventional superconductors have been discovered to exhibit evidence of spontaneous time-reversal symmetry breaking (TRSB) below their critical temperatures $T_c$~\cite{Ghosh2EA20,Henrik2022}. This is evidenced, for example, by enhanced muon-spin relaxation ($\mu$SR) and polar Kerr effect measurements reporting a change in the optical polar Kerr angle below $T_c$. This points to internal magnetic fields spontaneously generated by the superconducting state itself. A natural explanation for the occurrence of such fields would be either in terms of a nonunitary triplet pairing state or multi-component superconducting condensates entering a TRSB by a complex superposition of the order. For superconducting instabilities condensing in two-dimensional (2D) irreducible representations of the associated crystal point group, the latter interpretation appears particularly appealing. Complex superpositions of two symmetry-distinct order parameters may exhibit persistent supercurrents at material edges, dislocations, or around various defect sites. The multi-component scenario for the origin of TRSB below $T_c$ has been extensively discussed for e.g. Sr$_2$RuO$_4$~\cite{MaenoEA94, MackenzieEA17}, and heavy-fermion compounds UPt$_3$~\cite{StewartEA84, JoyntEA02}, UTe$_2$~\cite{RanEA19, AokiEA22}, PrOs$_4$Sb$_{12}$~\cite{BauerEA02, AokiEA03}, and also for Ba$_{1-x}$K$_x$Fe$_2$As$_2$~\cite{RotterEA08, BokerEA17}, LaNiC$_2$~\cite{HillierEA09}, and more recently additionally for the kagome superconductors $A$V$_3$Sb$_5$ ($A$: K, Rb, Cs)~\cite{Guguchia2022Tunable,Mielke2022Time-reversal,Romer2022}. At present, however, the status of the precise superconducting ground state remains controversial for several of these materials (e.g. Sr$_2$RuO$_4$ and UTe$_2$), partly due to specific heat data featuring only a single transition that does not split under uniaxial strain.

 Thus, it is important to pursue further possibilities for the origin of TRSB generated inside the superconducting phase. In particular, one might ask what mechanisms exist for TRSB for superconducting condensates composed of a single component order parameter? The answer to this question necessarily involves the presence of spatial inhomogeneities.
For example, it is well-known that nonmagnetic impurities can generate magnetic moments on nearby sites below $T_c$~\cite{Tsuchiura2001,ZWang2002,Zhu2002,Chen2004,Andersen2007,Harter2007,Andersen2007,Andersen2010,Schmid_2010,Gastiasoro2013,Gastiasoro2014,Gastiasoro2015,Martiny2015,Gastiasoro2016,Martiny2019}  in the presence of electronic correlations. In addition, one can envision disorder-induced slowing-down of magnetic fluctuations of preexisting magnetic impurities, thereby lowering the characteristic fluctuation frequencies into
 the muon time window. Another possibility for TRSB includes the generation of localized orbital loops of supercurrents. The latter is well-known to arise near nonmagnetic disorder sites in complex TRSB multi-component condensates~\cite{Lee2009,Garaud2014,Maiti2015,Lin2016,Silaev2017,GaraudPRL,Benfenati,Merce2022}, but was recently shown to be also present in the strong disorder limit of single-component superconductors~\cite{Li2021,Clara2022}. Likewise, one might expect dislocations and grain boundaries to similarly operate as seeds of localized loop currents. Recently, the latter scenario was proposed by Willa {\it et al.,}\cite{WillaEA21} as a possible explanation for reconciling the existence of TRSB and a single specific heat transition in Sr$_2$RuO$_4$. Indeed, for this material it is well-known that dislocations are prevalent in many samples~\cite{Ying2013}.

Here, motivated by the above-mentioned developments we perform a detailed theoretical study of the possibility of generation of localized supercurrents by dislocations in single-component order parameter superconductors. For concreteness, we focus on a one-band model with nearest neighbor attractive interactions.  This model supports either $B_{1g}$ ($d$-wave) or $A_{1g}$ (extended $s$-wave) spin-singlet superconductivity in the homogeneous case, depending on filling. In addition, there exists a coexistence regime where the complex combination $s+id$ is favored. We start from a single-band picture without any material-specific details, except the assumption that the orbital states are of $d_{x^2-y^2}$ symmetry. To obtain the modifications of the hopping integrals due to a dislocation, we remove a line of $l$ atoms and simulate the equilibrium positions assuming parabolic potentials between the atoms in a molecular dynamics setup. The hopping integrals are then the expectation value of the kinetic energy in atomic-like $d_{x^2-y^2}$ orbitals separated by the 
distances determined in this manner.
From selfconsistent real-space studies and associated calculations of the free energy, we find that the dislocations favor the local generation of complex pair potential order in a rather broad range of parameters and electron concentrations. This implies that TRSB dislocation-bound orbital currents are spontaneously stabilized. Our results highlight the role of spatial inhomogeneity in generating TRSB and associated magnetic signals within the superconducting phase.

\section{Model and Method}

\subsection{Dislocations and lattice relaxation}

\begin{figure}[t]
\centering
\includegraphics[angle=0,width=0.99\linewidth]{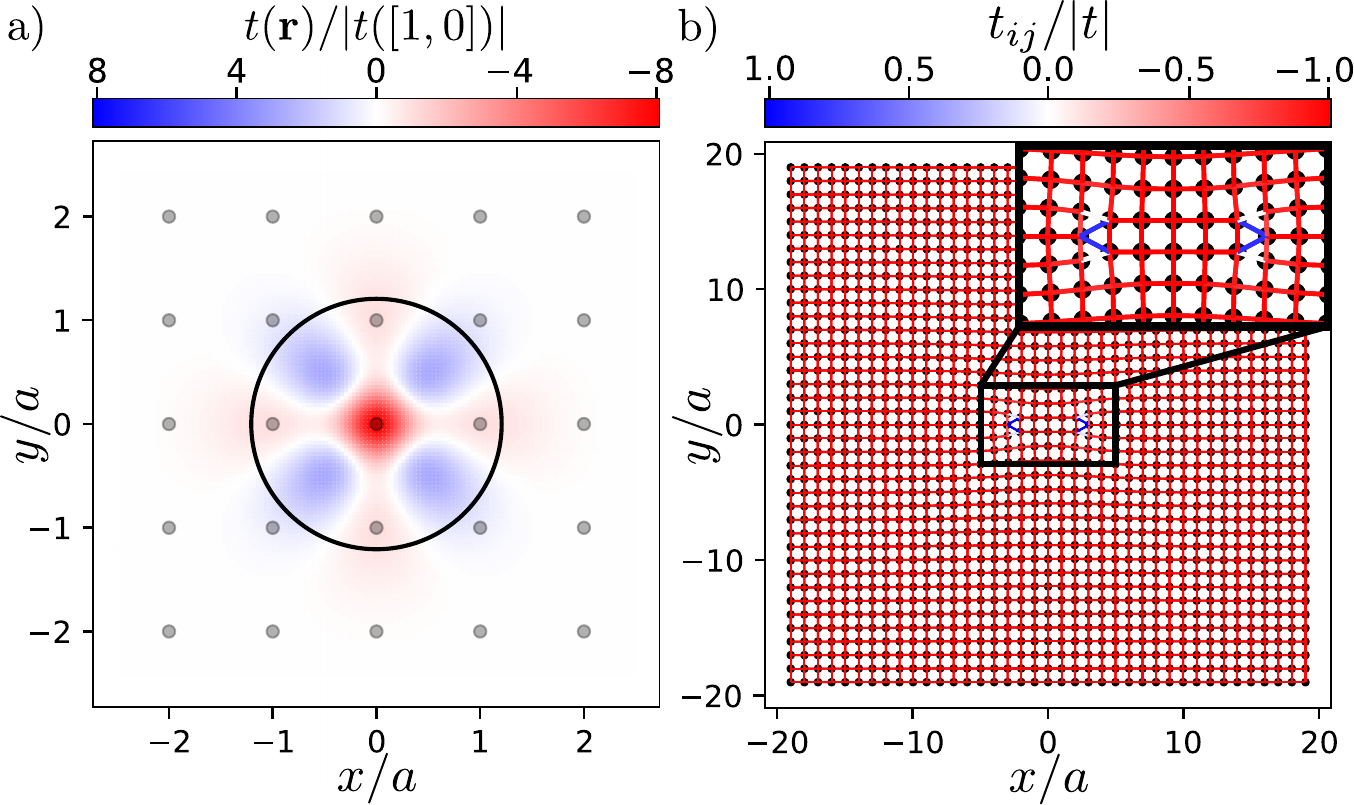}
\caption{(a)\label{fig:1} Hopping integral as function of connection vector $\mathbf r$ between sites. Black circle shows $r_{\mathrm{cut}} = (1 + \sqrt{2})/2$ defining the maximum distance allowed for NN bonds. Grey dots indicate site positions of the homogeneous lattice. (b) An example of real space points (black dots) describing a $l=5$ dislocation together with the value of the hopping integral (colored bonds). The inset shows a zoom-in of the region close to the dislocation.}
\end{figure}

The normal state Hamiltonian of a system of real space points with hopping matrix elements $t_{ij}$ is given by
\begin{equation}
H_0 = \sum_{i,j,\sigma} (t_{ij} - \mu \delta_{ij}) c^{\dagger}_{i\sigma}c_{j\sigma},
\end{equation}
where the operator $c^{\dagger}_{i\sigma}$ creates an electron at real-space point $i$ with spin $\sigma$.
For a homogeneous system, we set  $t_{ij} = t = -1$ if $i$ and $j$ are nearest neighbors (NN).
In the presence of a dislocation, the displacements of lattice points modifies the hopping integrals which we
capture in the following way. Given the distance vector $\mathbf{r}_{ij}=(x_{ij},y_{ij})$, we calculate the hopping from
the expectation value $t(\mathbf{r})$ of the kinetic energy $-1/(2m)\mathbf\nabla^2$ for overlapping atomic $d_{x^2-y^2}$ wavefunctions. A map of the hopping integral is shown in Fig.~\ref{fig:1}(a) where we used the value $|t([1,0])|=1$ to fix the effective mass $m$. Black circle shows $r_{\mathrm{cut}} = (1 + \sqrt{2})/2$ defining the maximum distance allowed for NN bonds. Hoppings for  $|\mathbf{r}_{ij}|>r_{\mathrm{cut}}$ are simply set to zero. Note that the implementation of a dislocation reduces the total number of sites in an $N \times N$ square lattice to $N^2 - l$.

The positions of the lattice points are obtained using a molecular dynamics simulation
with periodic boundary conditions where the sites of a square lattice are bound to NN and NNN sites
via unstretched strings of equal strength.
Next, a finite number $l$ of lattice sites along a line are removed to form a dislocation pair and bonds to the ``new'' NN and NNN along that line are established.
The relaxation under the harmonic potential with periodic boundary conditions is then done to find all positions $\mathbf{r}_i$ of the lattice points.
This procedure yields symmetric positions close to the dislocation and reproduces the strain expected from continuum theory far away from the dislocation.
 An example of the lattice positions (black dots) for $l=5$ is shown in Fig.~\ref{fig:1}(b) where the value of the hopping integrals are indicated by the color of the corresponding bond.

\subsection{Selfconsistent solutions of superconductivity}

\noindent The study presented here is based on the following one-band BCS mean-field Hamiltonian 
\begin{align}
H & =
H_0
- \sum_{\langle i,j \rangle} (\Delta_{ij} (c_{i\downarrow}^\dagger c_{j\uparrow}^{\dagger} - c_{i\uparrow}^\dagger c_{j\downarrow}^{\dagger}) + \mbox{H.c.} ) , \label{H}
\end{align}
where $\Delta_{ij} = V(\langle c_{j\uparrow} c_{i\downarrow}\rangle - \langle c_{j\downarrow} c_{i\uparrow} \rangle)$ is the attractive ($V>0$) singlet pairing on NN bonds also defined by $|\mathbf{r}_{ij}| < r_{cut}$.
The corresponding Bogoliubov-de Gennes (BdG) equations are solved selfconsistently for systems of size $N \times N$ with a dislocation of length $l$ as described above.
The convergence criterion is set to $|\mathcal{H}^s_{\alpha\beta} - \mathcal{H}^{s-1}_{\alpha\beta}| < 10^{-8}\, |t| \,\, \forall \,\, \alpha,\beta$, where $\mathcal{H}$ is the BdG Hamiltonian and $s$ denotes the iteration counter.
All results are computed with open boundary conditions (OBC) to avoid super-cell interference effects and we set $V = 0.75\,|t|$ and $N\times N = 39\times39$ unless otherwise specified. Additionally, the average electron density will be fixed by adjusting the chemical potential, $\mu$, in the selfconsistent iterations. Results presented here are obtained for the hole-doped region of the phase diagram ($\langle n\rangle <1$), but applies also to the electron-doped region due to particle-hole symmetry at half-filling ($\mu = 0$) with the symmetry $\langle n\rangle \rightarrow 2-\langle n\rangle$.

To study the possibility of spontaneous TRSB induced by the atomic dislocations and manifested by generation of orbital currents, the bond-current densities of all converged results are computed as
\begin{align}
\langle j_{ij} \rangle &= -i\frac{e}{\hbar a^2} \sum_{\sigma} t_{ij} \langle \cd_{i\sigma} c_{j\sigma} - \cd_{j\sigma}c_{i\sigma} \rangle \\\nonumber
&= 2\frac{e}{\hbar a^2} \sum_n t_{ij} \mathrm{Im} [u^*_{in\downarrow}u_{jn\downarrow} f(E_n) - v^*_{in\uparrow}v_{jn\uparrow}f(E_n)],
\end{align}
where $e$ denotes the electron charge, $u_{in\downarrow},\, v_{in\uparrow}$ are the usual eigenvector components from the Bogoliubov transformation and $f(E_n)$ is the Fermi-Dirac distribution function.  All results are obtained with the temperature fixed at $T = 0.001\,|t|$. The selfconsistency ensures current conservation as confirmed by incoming and outgoing currents being equal in magnitude at all lattice sites~\cite{barash}. We project the bond-current densities to local densities using
\begin{align}
\mathbf{j}_i = \frac{1}{2} \sum_{\langle i, j \rangle} \frac{\langle j_{ij} \rangle}{|\mathbf{r}_{ij}|} \begin{pmatrix} x_{ij} \\ y_{ij} \end{pmatrix}.
\end{align}

\subsection{Free energy calculation}

Introducing spatial inhomogeneities in the selfconsistent solution of the model vastly extends the parameter space in which the free energy should be minimized. As the selfconsistency only ensures results to be at stationary points in the free energy landscape, an enhancement of parameter space also increases the risk of obtaining converged results which do not represent the global minimum. Thus, to ensure that the spontaneous TRSB and generation of local loop currents are indeed energetically favorable, we compare the free energy of converged configurations with and without TRSB. The latter is obtained by restricting the superconducting order parameters to take real values at all bonds throughout the system. In this context, it should be noted that while Eq.~\eqref{H} is written in the grand canonical ensemble we compare solutions with fixed particle number. Thus, the relevant quantity to consider is the free energy rather than the grand potential i.e., $F = \Omega + \mu\hat{N}$, where $\Omega$ is the grand potential and $\hat{N} = \sum_{i\sigma} \langle \cd_{i\sigma}c_{i\sigma} \rangle$. In the zero temperature limit, this can be obtained as $F = \langle H \rangle + \mu\hat{N}$ yielding simply
\begin{align}
    F &= \sum_{\langle i,j \rangle, \,\sigma} t_{ij}\langle \cd_{i\sigma}c_{j\sigma}\rangle -\sum_{\langle i,j \rangle} \frac{|\Delta_{ij}|^2}{V}\,, \label{F}
\end{align}
where the expectation value is calculated using the eigenvalues $E_n$ and eigenvector components  $u_{in\downarrow},\, v_{in\uparrow}$ of the Hamiltonian Eq.~\eqref{H}.
Note that the constant term originating from the mean-field decoupling of the interactions which is not included in Eq.~\eqref{H} cancels the contribution from the expectation value of one of the superconducting pairing terms.

\section{Results}

\subsection{Homogeneous phase diagram}

\begin{figure}[tb]
\centering
\includegraphics[angle=0,width=\linewidth]{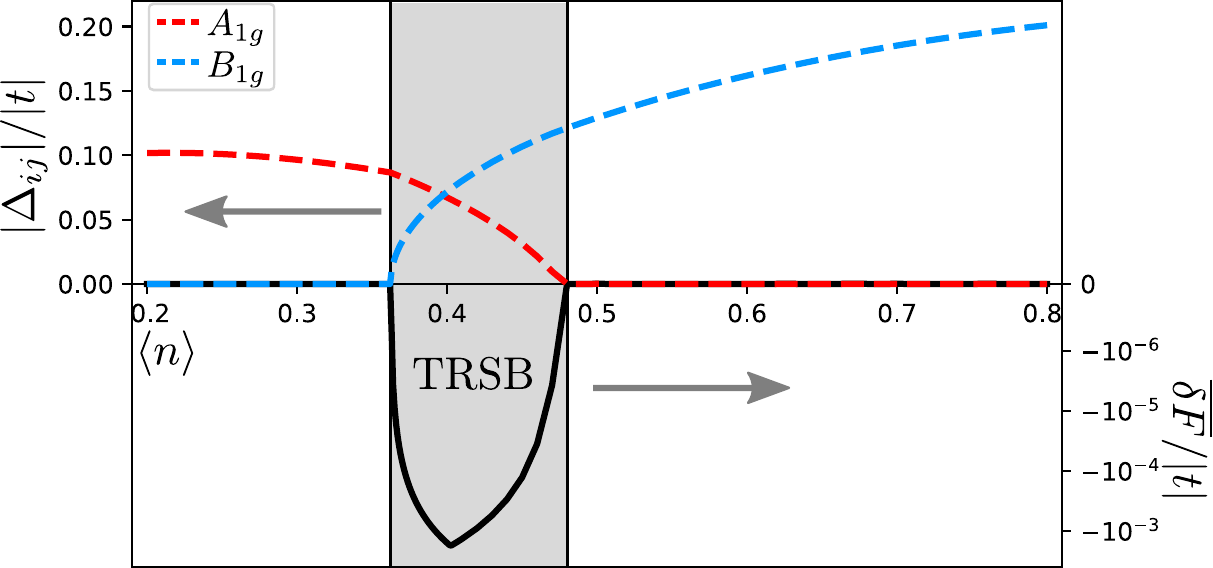}
\caption{Phase diagram showing on the left axis the amplitude of the homogeneous $A_{1g}$ (extended $s$-wave, red line) and $B_{1g}$ ($d_{x^2-y^2}$, blue line) order parameters as a function of electron density. The free energy difference per site, $\overline{\delta F} = (F_{\Delta \in \mathbb{C}} - F_{\Delta \in \mathbb{R}})/N^2$, is displayed on the right axis (black line). Results are computed on a $201\times201$ square lattice with periodic boundary conditions.\label{fig:2}}
\end{figure}

To establish a baseline for the results obtained for systems with dislocation defects, the phase diagram for the homogeneous system as a function of electron density is shown in Fig.~\ref{fig:2}. The magnitude of the $A_{1g}$ (extended $s$-wave, red line) and $B_{1g}$ ($d_{x^2-y^2}$, blue line) order parameters are shown on the left axis while the site-averaged free energy difference, $\overline{\delta F} = (F_{\Delta \in \mathbb{C}} - F_{\Delta \in \mathbb{R}})/N^2$, is indicated on the right axis (black line). Here $F$ is computed according to Eq.~(\ref{F}) and the subscript indicates whether $\Delta_{ij}$ are free to take on any value ($\Delta \in \mathbb{C}$) or they are restricted to real values preventing currents from developing ($\Delta \in \mathbb{R}$).
For electron densities close to half-filling, $\langle n \rangle = 1$, the $A_{1g}$ solution is strongly disfavored due to the near-alignment of the Fermi surface and line nodes, and therefore the selfconsistent solution exhibits $B_{1g}$ symmetry. On the contrary, for small Fermi pockets ($\langle n \rangle < 0.36$), a full gap can develop in the $A_{1g}$ channel which is more favorable than the $B_{1g}$ channel hosting four point nodes at the Fermi surface for all electron densities~\cite{Kreisel2017,Romer2015,Merce2022}.

In the intermediate region ($0.36 < \langle n \rangle < 0.48$) a complex superposition of the two channels is favored, yielding a TRSB solution of the form $A_{1g} \pm iB_{1g}$. A possible nematic solution $A_{1g} \pm B_{1g}$ exhibits line nodes at $k_x = \pm \pi/2$, while the TRSB solution reduces the line nodes to point nodes at $k_x = \pm k_y = \pm \pi/2$, resulting in a gain in condensation energy. This energy gain is reflected by the free energy difference shown in black in Fig.~\ref{fig:2}. Note by the definition stated above, that a negative value of $\overline{\delta F}$ signifies a TRSB ground state. Pure $A_{1g}/B_{1g}$ regions have $\overline{\delta F} = 0$, since the two solutions ($\Delta \in \mathbb{C}$ and $\Delta \in \mathbb{R}$) can only differ by a change
in the global superconducting phase.

\subsection{Loop currents and pairing modulations}

Several earlier studies investigated the emergence of orbital currents induced by non-magnetic defects in TRSB superconductors employing both phenomenological Ginzburg-Landau approaches and microscopic mean-field methods~\cite{Maiti2015,Graf2000,Lee2009,Seibold2015,Lin2016,Li2021,Clara2022}. In particular, the studies reveal that superconductors with $A_{1g} + iB_{1g}$ pairing symmetries generate spontaneous supercurrents in the vicinity of various types of non-magnetic potential scatterers, including point-like and spatially extended impurities~\cite{Clara2022,Maiti2015}, system boundaries~\cite{Lee2009} as well as defects exhibiting non-trivial spatial structures~\cite{Lin2016}. While the dislocations introduced in this work do not act as simple potential scatterers, it is reasonable to assume that the spatial inhomogeneities also generate orbital localized currents in the $A_{1g} + iB_{1g}$ phase due to local fluctuations in the superconducting order parameter analogous to the results presented in previous studies~\cite{Maiti2015,Graf2000,Lee2009,Seibold2015,Lin2016,Li2021,Clara2022}. Indeed, the computed selfconsistent results for systems in the homogeneous TRSB density range including a dislocation in the center region confirm the formation of local loop currents surrounding the defect. We note that, while the specific current pattern and magnitude depend on the dislocation length, the formation of currents is robust for all dislocations studied here. 

\begin{figure}[tb]
\centering
\includegraphics[angle=0,width=\linewidth]{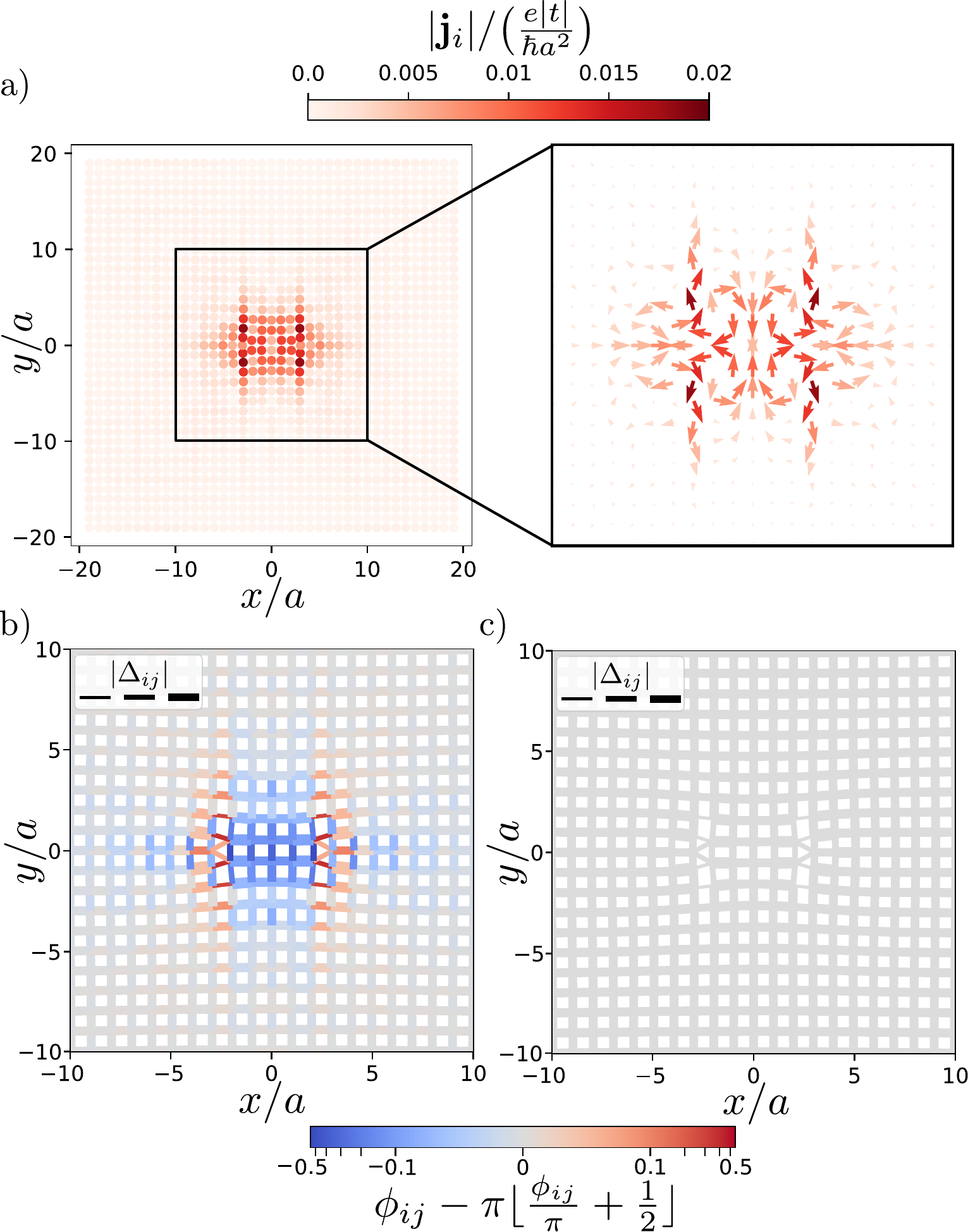}
\caption{(a) Site-resolved magnitude of the current density. Localized loop currents have developed near the dislocation with a small, long-range, tail required by current conservation. Zoom-in show the local current direction indicated by arrows. Arrow lengths have been re-scaled for visual clarity. (b,c) Modulations of $\Delta_{ij}$ for the converged solution with (b) and without (c) currents. Bond thickness indicate $|\Delta_{ij}|$ and bond color display modulations of the bond phase, $\phi_{ij}$. Here $\lfloor . \rfloor$ is the floor-function such that the nearest multiple of $\pi$ is subtracted and only the deviations in $\phi_{ij}$ away from pure $B_{1g}$ structure is shown. All results are computed with an $l = 5$ dislocation at $\langle n \rangle = 0.64$. \label{fig:3}}
\end{figure}

While the generation of defect-induced supercurrents in the $A_{1g} + iB_{1g}$ phase might be expected, the emergence of currents in the pure $A_{1g}$ or $B_{1g}$ phases  would require spontaneous TRSB occurring solely due to the lattice inhomogeneity associated with the presence of a dislocation. Surprisingly, we find that this scenario is indeed realized for a range of densities well within the homogeneous $B_{1g}$ phase. This is unlike the property of single impurities which are unable to seed orbital currents in this region. Figure~\ref{fig:3}(a) displays the induced orbital currents for result obtained at $\langle n \rangle = 0.64$ and $l = 5$. The supercurrents are predominantly localized in the region extending $\sim 5$ lattice sites away from the dislocation, but current conservation requires a small, yet finite, long-range tail, see left panel of Fig.~\ref{fig:3}(a). The lack of corner/edge currents emphasises that the near-homogeneous regions remain in the $B_{1g}$ phase and, as such, the apparent breaking of TRS is a purely local and dislocation-induced effect. The zoom-in shown in the right panel displays the current direction, demonstrating the orbital loops in the supercurrent pattern. The magnitudes are indicated by arrow color, while the arrow lengths have been re-scaled for visual clarity.

Since supercurrents arise from modulations in the phase and amplitude of the order parameter, we display the bond order parameters corresponding to Fig.~\ref{fig:3}(a) in Fig.~\ref{fig:3}(b). For comparison, Fig.~\ref{fig:3}(c) shows the converged result for identical parameters with the $\Delta_{ij} \in \mathbb{R}$ restriction applied (i.e. no currents allowed). The line thickness indicate the amplitudes of the bond pairings while the color displays the phase-deviation away from $\phi_{ij} = 0, \pm \pi$, i.e. the deviation of the superconducting phase away from a pure $B_{1g}$ phase with e.g. $\phi_{ij} = 0$ on $x$-bonds and $\phi_{ij} = \pi$ on $y$-bonds. The restriction applied to the result shown in Fig.~\ref{fig:3}(c) obviously hinders phase modulations and the spatial inhomogeneity associated with the dislocation mainly serve as a pair-breaking effect as evident from the smaller gap amplitude on the bonds near the ends. In contrast, the current-carrying solution shown in Fig.~\ref{fig:3}(b) features clear deviations from the pure $B_{1g}$ phase up to $18 \%$ alongside the amplitude fluctuations. Explicit comparison of the free energy difference per site in this particular case yields indeed a negative $\overline{\delta F} = -2.4\cdot10^{-6} |t|$. This is much smaller than the typical reduction of the free energy when a homogeneous $A_{1g}+iB_{1g}$ state emerges; again a signature of the local nature of TRSB in this scenario.

\subsection{Dislocation phase diagram and re-emerging supercurrents}

\begin{figure*}[bt]
\centering
\includegraphics[angle=0,width=0.99\linewidth]{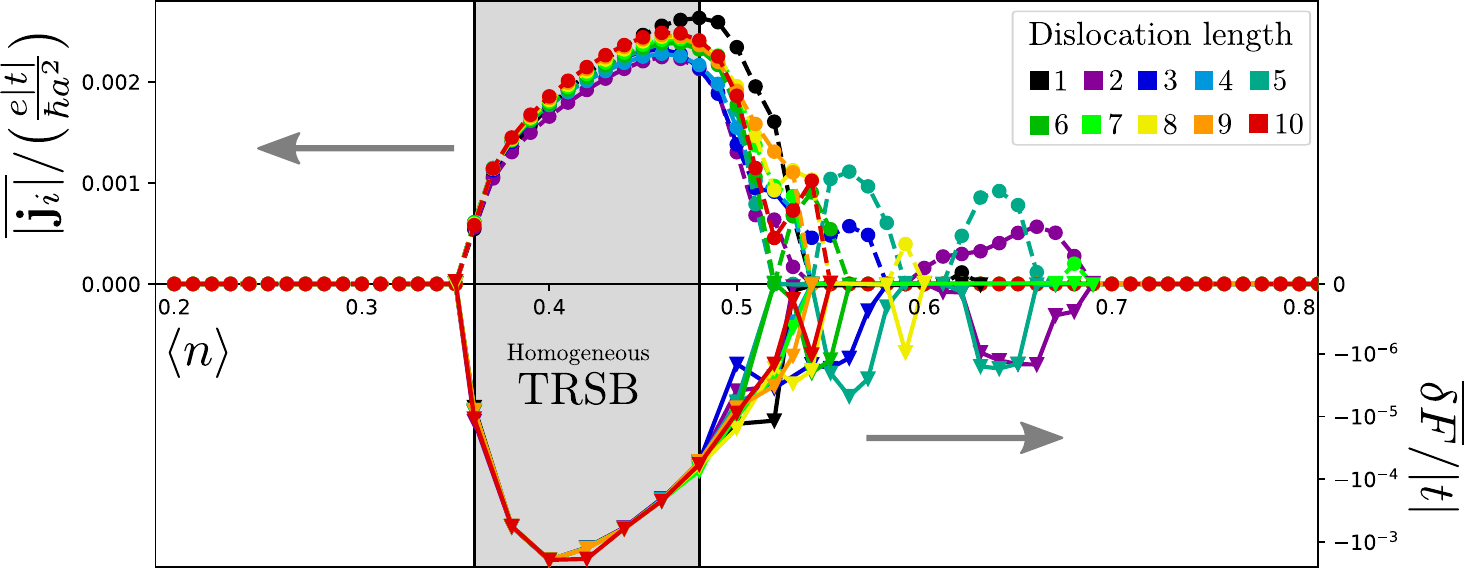}
\caption{\textit{Left axis:} Site-averaged local current density magnitude (dashed lines, circle markers). A main TRSB dome extends from the homogeneous TRSB region (marked in grey) with re-emerging domes extending to $\langle n \rangle \sim 0.7$. \textit{Right axis:} Site-averaged free energy gain (solid lines, triangle markers) associated with the spontaneous TRSB. All TRSB solutions are energetically favored or, in rare cases, degenerate with the time-reversal symmetric solution. Colors indicate different dislocation lengths, $l$. Note that the total number of sites used in the averages is given by $N^2 - l$.\label{fig:4}}
\end{figure*}

Given the rather unexpected finding of energetically favorable loop currents far inside the homogeneous $B_{1g}$ phase, it is desirable to perform a more extensive study of this phenomenon as a function of dislocation length and electron density. Thus, in Fig.~\ref{fig:4} the left axis shows site-averaged supercurrent (dashed lines, circle markers) for dislocation lengths ranging from $l = 1$ to $l = 10$ as a function of density. As a direct reference to the phase diagram in Fig.~\ref{fig:2}, the homogeneous TRSB region has been marked in grey. It is evident that all dislocation lengths induce currents within this region as anticipated from the previous studies discussed above. We note, however, that the dome emerging from the homogeneous TRSB region is significantly enhanced with the upper phase boundary shifted from $\langle n \rangle = 0.48$ to $\langle n \rangle \sim 0.53$. 

More interestingly, the generation of loop currents is clearly extended far beyond this main dome with re-emerging TRSB domes ranging all the way to $\langle n \rangle \sim 0.7$. While the density ranges and current strengths of the TRSB domes varies for different dislocation lengths, the multi-dome feature is universal across all systems investigated here, where some show an additional shoulder at the edge of the main dome (e.g. $l = 9$, marked in orange) and other show clearly distinguishable domes (e.g. $l = 2$, marked in purple, or $l=5$, marked in cyan).

Figure~\ref{fig:4} also shows the free energy gain (solid lines, triangle markers) associated with the generation of loop currents on the right axis verifying that the TRSB solution is energetically preferred with the exception of a few points with degenerate time-reversal symmetric and TRSB solutions down to the convergence criterion, see e.g. $l = 7$ (light green) at $\langle n \rangle = 0.68$. Note that the individual contributions from the two terms in Eq.~\eqref{F} varies between loss and gain for different parameters and only the total free energy difference reveal a consistent gain across all TRSB solutions.

\section{Discussion and conclusions}

It is relevant to highlight a fundamental difference between the spatial inhomogeneity introduced by dislocation defects and multiple impurities acting as dopants. Refs.~\cite{Li2021} and \cite{Clara2022} both found qualitatively similar results for strongly disordered systems ($>20\%$ dopant atoms) where spontaneous loop currents are formed well within the homogeneous $B_{1g}$ phase. However, such strong disorder breaks all spatial symmetries and, as argued in Ref.~\cite{Clara2022}, unavoidably introduce finite components of the other pairing symmetry channels. In contrast, the dislocation defects studied in this work preserves a global $C_2$-symmetry ($C_4$-symmetry for $l = 1$) around $\mathbf{r} = (0,0)$ and, from a naive perspective, it is reasonable to assume that disorder with a higher symmetry should have less impact on the preferred symmetry of $\Delta_{ij}$. However, as evident from the results presented here, the disorder does not need to be strong from a symmetry-breaking perspective to obtain TRSB ground states. 

This is quite important, because many unconventional superconducting crystals are rather clean by many standard criteria, yet undoubtedly contain significant concentrations of such linear defects oriented by strain, which may be preferentially along high symmetry axes. For example, the highest quality Sr$_2$RuO$_4$ crystals are famously extremely clean; tiny amounts of nonmagnetic disorder (e.g. 0.15\% Ti$^{4+}$ substituting for Ru$^{4+}$) suppress $T_c$ to zero~\cite{Mackenzie1998,Kikugawa,KikugawaPRB}. Both Kerr and $\mu$SR measurements indicate a small TRSB signal at the superconducting transition, yet the ``standard" explanations in terms of two-component order parameters are seriously challenged by the absence of any splitting of the specific heat transition with or without strain~\cite{DeguchiEA04,LiEA19}, suggesting alternative explanations for the weak TRSB.  One such explanation was provided by Willa {\it et al.}\cite{WillaEA21}, who assumed a close proximity of two pair components, $d$- and $g$-wave such that a TRSB state was nucleated near an edge dislocation. Here we have demonstrated a similar scenario within a microscopic model, and also shown that the degeneracy need not be particularly close.

Additionally, apart from a few bonds located at the dislocation edges, all bonds exhibit near-homogeneous values of the hopping integrals leading to the expectation that a near-homogeneous $B_{1g}$ phase should develop. To examine this in detail, we performed a systematic investigation of the amount of lattice distortion necessary to reproduce TRSB domes within the homogeneous $B_{1g}$ phase and found within additional self-consistent calculations that even small modifications in the hopping integrals down to $5\%$ must be included to \textit{qualitatively} reproduce the results shown in Fig.~\ref{fig:4}. To \textit{quantitatively} reproduce the results, all relaxation effects must be included. Furthermore, we have also found that the particular density ranges and current magnitudes shown in Fig.~\ref{fig:4} varies somewhat with system size and boundary conditions (e.g. employing periodic boundaries). We stress, however, that the generic re-emergence of TRSB ground states within the homogeneous $B_{1g}$ phase persists for a vast range of investigated system sizes and boundary conditions. These combined  findings suggest that the weak long-range perturbations and currents induced by the dislocation are essential. In this context, it is worth mentioning that typical dislocations in crystals may involve many hundreds of atoms, which presumably enhances these long-range effects.

In summary, we have performed a theoretical study of the preferred superconducting order near dislocations of a microscopic model that incorporates two symmetry-distinct superconducting order parameters. It is found that dislocations tend to nucleate orbital currents and thereby locally break time-reversal symmetry spontaneously. This takes place even in regions of the phase diagram where the corresponding homogeneous superconducting condensate is single component order. This effect may be important for the understanding of experiments reporting TRSB at $T_c$ in unconventional superconductors.

\section*{Acknowledgements} We acknowledge useful discussions with C. Hicks and M. Pal. C.N.B. and A.K. acknowledge support by the Danish National Committee for Research Infrastructure (NUFI) through the ESS-Lighthouse Q-MAT. P.J.H. acknowledges support from NSF-DMR-1849751. M.R. acknowledges support from the Novo Nordisk Foundation grant NNF20OC0060019.

% \nocite{*}
 \bibliography{litlist}
\end{document}